\documentclass[11pt]{article}
\pdfoutput=1
\usepackage{graphicx}
\usepackage{epstopdf}
\AppendGraphicsExtensions{.tif}

\usepackage[english]{babel}
\usepackage[ansinew]{inputenc}
\usepackage{epsfig}
\usepackage{float}
\usepackage{fullpage}
\usepackage{wrapfig}
\usepackage{float}
\usepackage{amsfonts}
\usepackage{amssymb}
\usepackage{colortbl}
\usepackage{mathcomp}
\usepackage{gensymb}
\usepackage{amsmath}

\usepackage{setspace}

\begin{document}


\title{High accuracy decoding of dynamical motion from a large retinal population}

\author{Olivier Marre$^{1,2}$, Vicente Botella-Soler$^{3}$, Kristina D. Simmons$^{4}$, \\
Thierry Mora$^{5}$, Ga\v{s}per Tka\v{c}ik $^{3}$ and Michael J. Berry II$^{1}$}

\maketitle

\small $^{1}$Department of Molecular Biology and Neuroscience Institute, Princeton University ; $^{2}$ Institut de la Vision, INSERM UMRS 968, UPMC UM 80, CNRS UMR 7210, Paris ; $^{3}$ IST Austria ; $^{4}$ Department of Psychology, University of Pennsylvania ; $^{5}$ Laboratoire de Physique Statistique, Ecole Normale Superieure, Paris \\
\\

\abstract{ 
Motion tracking is a challenge the visual system has to solve by reading out the retinal population. Here we recorded a large population of ganglion cells in a dense patch of salamander and guinea pig retinas while displaying a bar moving diffusively. We show that the bar position can be reconstructed from retinal activity with a precision in the hyperacuity regime using a linear decoder acting on 100$+$ cells. The classical view would have suggested that the firing rates of the cells form a moving hill of activity tracking the bar's position. Instead, we found that ganglion cells fired sparsely over an area much larger than predicted by their receptive fields, so that the neural image did not track the bar. This highly redundant organization allows for diverse collections of ganglion cells to represent high-accuracy motion information in a form easily read out by downstream neural circuits. 
}

\newpage

\section{Introduction}

Our current understanding of how sensory neurons collectively encode information about the environment is limited. Being able to ``read out'' this information from neural ensemble activity is a major challenge in neuroscience. Discriminating among a discrete set of stimuli based on their sensory responses has been attempted in brain areas including the retina \cite{FrechetteEJ, Schwartz2012}, sensory cortex \cite{Oram1998, Kay2008}, and motor cortex \cite{Georgopoulos}. Some studies have attempted the more difficult task of reconstructing a time varying signal from neural activity \cite{Wilson1993, Zhang1998, Pasley2012}. Being able to reconstruct a dynamical stimulus from the neural activity would greatly improve our understanding of the neural code, but there have been only a few attempts in the visual system \cite{Bialek1991, Warland1997,pillow2008}. 

The retina is an ideal circuit in which to attempt to decode a dynamical stimulus because recordings from a large, diverse, complete population of ganglion cells---the retinal output---have recently become possible \cite{Field2010,Method252}. In contrast to cortical recordings, it is possible to drastically reduce the proportion of  ``hidden variables'' in the network, i.e. unrecorded neurons that could carry relevant information.  Furthermore, since the retina encodes all visual information available to the brain, the decoding performance of a complete retinal population can be rigorously compared to behavioral performance for an equivalent task \cite{Segev2007, Jacobs2009}.

Motion tracking is of major ecological relevance. Amphibians can capture small moving prey, like flies, at a variety of speeds and distances from their body \cite{Deban1997,Roth1987,Llinas2000}, implying that the retina must track such motion accurately. Furthermore, humans can use fixational eye movements to discriminate stimuli separated by roughly two cone photoreceptors, which is highly challenging without an accurate retinal representation of image movement \cite{Pitkow2007}.

Yet a reconstruction of the position of an object moving randomly from the activity of sensory neurons has not been achieved in the vertebrate visual system. 
The classical view suggests that ganglion cells will signal the position of the bar when it is at the peak of their receptive field, making this reconstruction easy by tracking the peak firing rate in the retinal map \cite{Rodieck1965,Leonardo2013}. However, the non-linear computations performed by the retinal network seem to make this picture more complex than intially thought: for example, a synchronous peak in the activity could signal sudden changes of speed \cite{Thiel2007} or a sudden reversal of motion \cite{Schwartz2007}. So it is unclear how a complex trajectory can be decoded from the retinal output, and which neurons are most useful for that purpose. 

In order to study how neural populations encode dynamic motion, we recorded from a large fraction of the ganglion cells in a patch of the retina while stimulating with a randomly moving dark bar. 
We show that the entire trajectory of the bar can be estimated by a linear decoder with a precision better than the average spacing between cones using groups of 100$+$ ganglion cells. 
Our analysis also revealed an unexpected structure to the population code.  Instead of representing the object's location with a ``hill'' of neural activity, the population activity was sparse and broadly distributed in space. The retina thus used a distributed and redundant code, in which several disjoint groups could be read-out to reconstruct the stimulus with high precision. 

\section{Results}

\subsection{Population recording and linear decoding}

We used a large multi-electrode array with 252 electrodes to record the responses of ganglion cells in the salamander and guinea pig retinas (see Methods), while presenting a randomly moving bar (fig.~1A,B). The density of the electrodes allowed us to record a large fraction of the ganglion cells in the retina patch covered by the array \cite{Method252}. The neurons were recorded over a compact region and had highly overlapping receptive fields. Up to 189 cells were recorded simultaneously (see example spike rasters in fig.~1C) over several hours. 
The stimulus was a dark bar (width = 100 $\mu$m) on a grey background moving diffusively over the photoreceptor layer. The trajectory was a random walk with a restoring force to keep the bar close to the array (fig.~1A, and Methods), spanning a region corresponding to roughly 10 degrees of visual angle.  

One of our primary motivations in choosing this stimulus ensemble was to select patterns of motion that are more complex than constant velocity punctuated by sudden discontinuities. Many previous studies of motion processing in the retina have used objects moving at a constant velocity, sometimes punctuated by sudden changes of velocity \cite{Berry1999,Thiel2007,Schwartz2007,Leonardo2013}. While such studies have contributed greatly to our understanding of retinal motion processing, they are not necessarily representative of the patterns of motion that an animal encounters in the natural environment. Obviously, not all objects in the world move at nearly constant velocity for long periods. In addition, we suspected that ganglion cell responses to complex motion might not be simply related to the manner in which they respond to constant velocity motion. Thus, we wanted a more general class of motion patterns. Building on the analogy to white noise stimuli, we selected a class of diffusive motion, which constitutes a broad ensemble of motion patterns without making highly specific choices about the set of trajectories. 

\begin{figure}[!]
\centering
\includegraphics[width=0.9\textwidth]{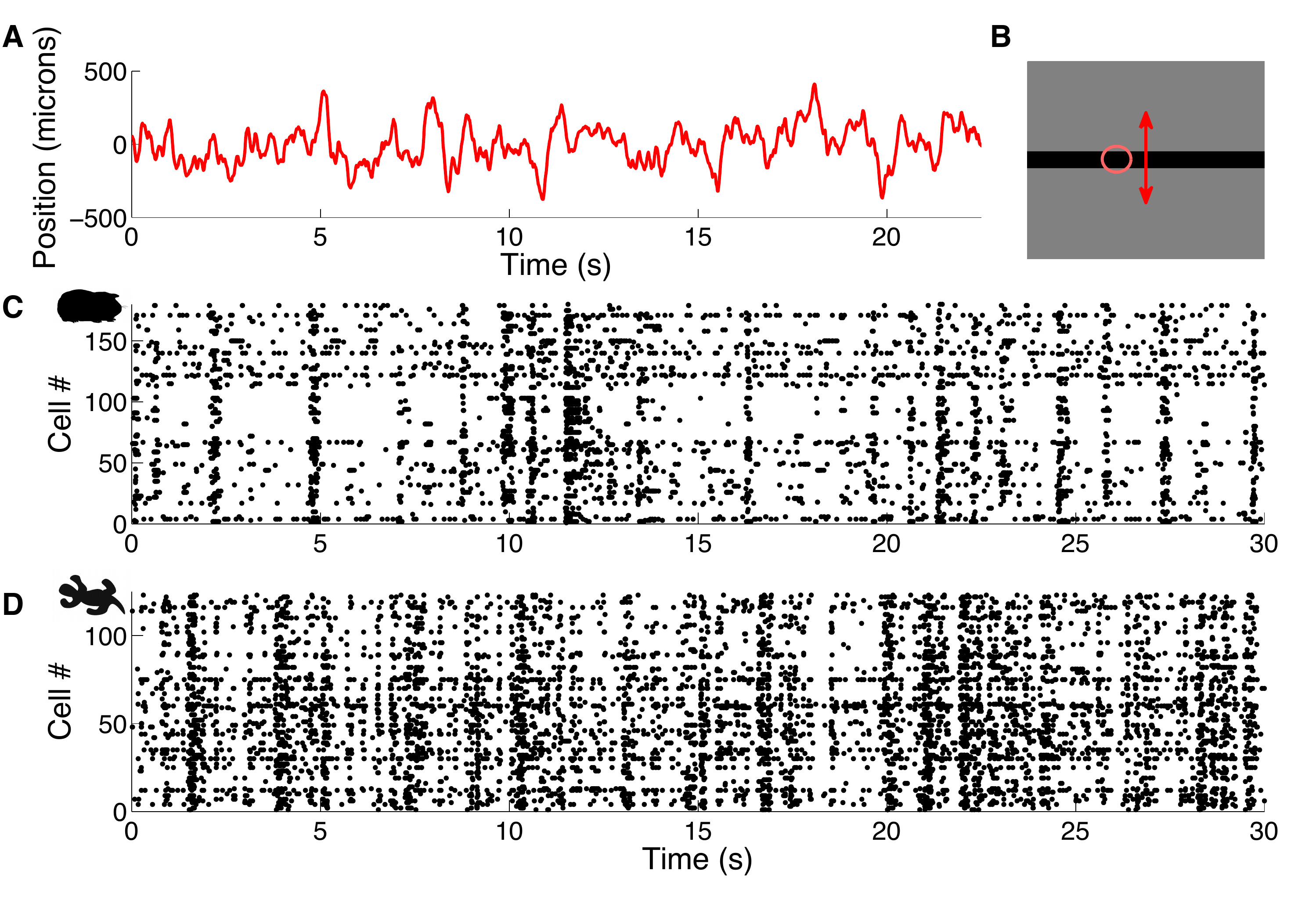}
\caption{ 
\textbf{Ganglion cell spike trains during random bar motion. } 
\textbf{A:} Position of the bar as a function of time. 
\textbf{B:} Example of one stimulus frame; motion is perpendicular to the bar (red arrow). Ellipse fitted to the spatial receptive field profile of one representative ganglion cell (pink). 
\textbf{C:} Spiking activity of 180 cells in the guinea-pig retina in response to a bar moving randomly with the trajectory shown in A. Each line corresponds to one cell, and the points represent spikes. The order of the cells along the y-axis is arbitrary. 
\textbf{D:} Spiking activity of 123 cells in the salamander retina responding to a bar moving randomly. Same convention as \textbf{C}. 
}
\label{Fig1}
\end{figure}

Under these stimulus conditions, the activity of ganglion cells was sparse and strongly modulated (fig.~1C and D). Our goal was to decode the position of the bar at all times from this retinal activity. We used a linear decoder that took as an input the spike trains of all the ganglion cells and predicted the position of the bar as a function of time \cite{SpikesBook}. A temporal filter is associated with each cell and added to the prediction each time the cell fires an action potential (fig.~2A). The filter shapes were fitted on a training fraction of the data ($2/3$) to minimize the squared error between the predicted and true trajectories  (see Methods for details). To test the decoder, the prediction was evaluated on the previously withheld test fraction of the data ($1/3$). 

With 140 cells recorded in a single experiment in the guinea pig retina, the prediction was very precise and followed closely the position of the bar (fig.~2B). The normalized cross-correlation (see Methods) between the two traces was CC=0.95 (corresponding to an error of $\sim$ 5$\mu$m, or 0.15 degree of visual angle) on the testing and the training dataset, which also indicated that the effects of over-fitting were minimal. Similar results were found in the salamander retina (with CC=0.9 in the testing set, CC=0.93 on the training dataset, fig.~2C). They were confirmed in $n=4$ retinas for the salamander, and $n=2$ for the guinea pig. Similar results were obtained when forcing the filters to be causal, by restricting their parameters to the range of -500 ms to 0 ms before a spike (CC=0.9). Thus, the activity of a large population of the ganglion cells contained enough information to reconstruct the position of an object moving randomly across the visual field with high precision. Furthermore, a simple readout mechanism - the linear decoder - was able to perform this reconstruction. 

How large a population of cells is necessary to decode the position of the bar so precisely? To address this question, we selected random subsets of cells, performed linear decoding on them, and quantified their decoding performance by measuring the cross-correlation between the real and the predicted trajectories. The decoding performance as a function of the number of selected ganglion cells grew rapidly for few cells and then slowed down for $\gtrsim$70 cells in both the guinea pig and the salamander retina (fig.~2D,E). Therefore, recording a large number of ganglion cells in the same area of the retina was essential to be able to track the trajectory of a moving object with high accuracy. 

\begin{figure}[!]
\centering
\includegraphics[width=0.7\textwidth]{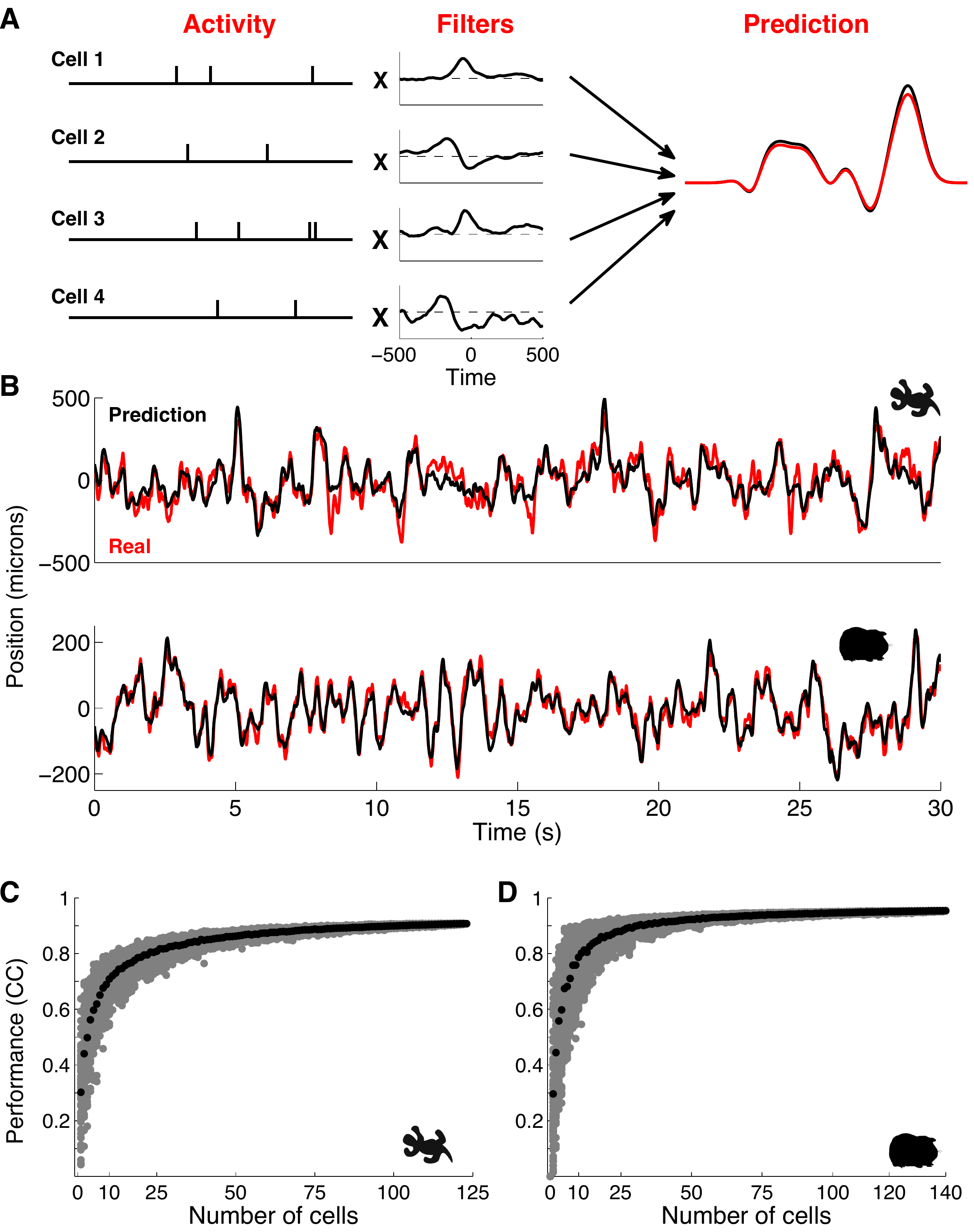}
\caption{ 
\textbf{High-accuracy reconstruction of the bar's trajectory. }
\textbf{A:} Schematic of the linear decoding method, here for 4 cells. A temporal filter is associated with each cell. Each time the cell spikes, its filter is added to the ongoing reconstruction at the time of the spike. The filters are optimized on part of the data to have the lowest reconstruction error and then tested on the rest of the data. 
\textbf{B:} Prediction of the bar's position (black) from the activity of 123 cells in the salamander retina versus the real trajectory (red). 
\textbf{C:} Prediction of the bar's position (black) from the activity of 178 cells in the guinea pig retina versus the real trajectory (red). 
\textbf{D, E:} Decoding performance plotted against the number of cells in the salamander (D) and guinea pig (E). Gray points correspond to random subsets of cells, black to the average performance. 
}
\label{Fig2}
\end{figure}

\subsection{Performance estimation}

While the high cross-correlation between the real and predicted bar trajectories demonstrates high performance, it is instructive to make this comparison along other dimensions. The capacity of the retinal circuit to track the position of a moving object is related to the geometry and signal-to-noise ratio of its sensors \cite{Ruderman1992}. We thus wanted to analyze more precisely the prediction errors and relate them, at least qualitatively, to the properties of the photoreceptor array.

This prediction performance is likely to depend on the speed of the bar. To quantify this dependence, we generated a new bar trajectory: a white noise trajectory low-pass filtered at 5 Hz (see Methods). This stimulus ensemble is desirable because a large band of frequencies are represented equally (fig. 3A; blue). This allowed us to explore how the retina represents both slow and fast fluctuations in the bar's position. We recorded the responses of 158 neurons in the salamander retina to this moving bar, performed the same linear decoding, and estimated the squared prediction error as a function of frequency (fig.~3A). The error was clearly frequency-dependent with a dip between 1 and 4 Hz. Since the photoreceptors are themselves low-pass filters \cite{Rieke2000}, the best decoding performance is necessarily bounded by their sensitivity spectrum \cite{Ruderman1992}. 

But the bar's motion also spanned a wider field than the array, so we hypothesized that errors would be larger for eccentric positions. We plotted the absolute difference between the real trajectory and the prediction against the position of the bar itself, for each time point (fig.~3B). As expected, the error was position-dependent with the minimal error found for the position that was at the middle of the bar's range of motion. Note that this dependence on position can have more than one explanation. It can come from an incomplete sampling of the side positions by the cell recorded, but also from the fact that, when minimizing the error, more weight is given to the center positions, since they are explored more often.  The error spectrum estimated above thus reflected an average between heterogeneous error levels taken at different positions of the bar. We thus aimed at separating simultaneously the effects of position and signal frequency on the prediction error. 

For that purpose, we designed an error measure that took these two aspects into account. For each frequency, we filtered the signal with a band-pass filter centered on that frequency, and then computed the variance separately for each position (see Methods). For proper normalization, we required that a weighted average of this position-frequency spectrum over the positions give back the power spectrum shown in (fig.~3A). At the middle of the bar's range of motion, the error spectrum had a soft dip between 1 and 4 Hz, a slightly broader range of frequencies than when averaging over all bar positions (fig.~3C). Over that range of frequencies, the error was significantly below the average spacing between cones in the salamander retina \cite{Sherry1998,Wu2004}, illustrated by the dashed line. To our knowledge, this is the first demonstration of high precision in decoding the time-varying trajectory of an object's motion in the vertebrate visual system. 

\begin{figure}[!]
\centering
\includegraphics[width=1\textwidth]{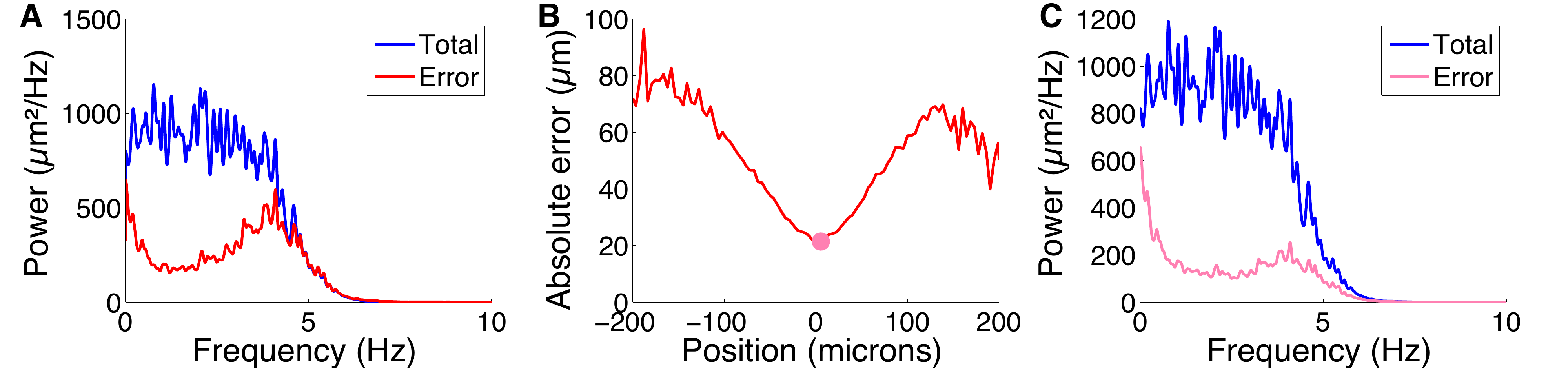}
\caption{ 
\textbf{The prediction error is in the hyperacuity range. }
\textbf{A:} Squared error as a function of frequency (red), compared to the power spectrum of the trajectory (blue). 
\textbf{B:} Root mean-squared error as a function of position (red) with the point of minimal error (pink circle). 
\textbf{C:} Error spectrum as a function of frequency (pink), for the position labeled in \textbf{B}. Power spectrum of the full trajectory (blue) ; error spectrum corresponding to the spacing between cone photoreceptors (dashed line). 
}
\label{Fig3}
\end{figure}

\subsection{Decoding based on the neural image}

Our naive expectation was that ganglion cells would fire when the bar moved in their receptive field center, such that the population had a peak of activity that travelled with the moving object \cite{Berry1999}. 
In this case, the moving object's position is represented well by the peak in the neural image. Following a sudden reversal of motion, there is an error in the location of the neural image \cite{Schwartz2007}. However, the spatial location represented by the synchronous burst of firing following motion reversal was shifted in the new direction of motion, so that it begins to ``correct'' for the retina's mistaken representation of the object position. As a result, a decoder based on the population vector could achieve good results for an object moving at constant velocity punctuated by sudden reversals \cite{Leonardo2013}. A related example comes from place cells in the hippocampus, where a simple population vector decoder can reconstruct the position of the animal with good resolution \cite{Wilson1993, Zhang1998}. 

Thus, we wanted to test if this previously used population vector decoding could be successful at predicting the bar's trajectory, and see if the retinal activity showed a moving hill that followed the bar's position. Compared to our previous method of linear decoding, it is interesting to note that population vector decoding corresponds to linear decoding with a temporal window of just a single time bin, and the weight of the filter being related to the value of the receptive field center. 

We first constructed the ``neural image'' as in previous studies, where it's peak corresponded well with the position of the moving bar \cite{Berry1999,Schwartz2007}.  The neural image is the spatial pattern of firing in the ganglion
cell population as a function of time. We calculated it by plotting the firing rate of the cells as a function of their receptive field position (see Methods). 
Despite the success of the neural image in tracking simpler object motion \cite{Berry1999,Leonardo2013}, we found that it did a poor job with our complex trajectory (fig.~4A, vs CC$=0.06$). Estimating an average of the position weighted by the firing rate obtained at each position did not improve the estimation of the real position (CC$=0.027$). One of the primary reasons for this poor performance was that neural activity was too sparse to continuously represent the moving object's location (fig.~4B).
We quantified the sparseness of a cell as in \cite{Berry1997}, by measuring the amount of time where the firing rate was above 5\% of its maximum value. On average, we obtained that it was above this threshold 11\% of the time (n$=$117 cells), which means that neurons remained inactive 89\% of the time. The population activity was also sparse: it was above 10\% of its maximum only 36\% of the time. 
Even outside the periods of complete silence, the location of the peak of neural activity was weakly correlated with the position of the bar (fig.~4A, CC$=0.06$). Thus, there was no clear moving hill of neural activity from which the position of the bar could be inferred. Instead, we observed sparse neural activity with no obvious spatial structure. 

\begin{figure}[!]
\centering
\includegraphics[width=0.9\textwidth]{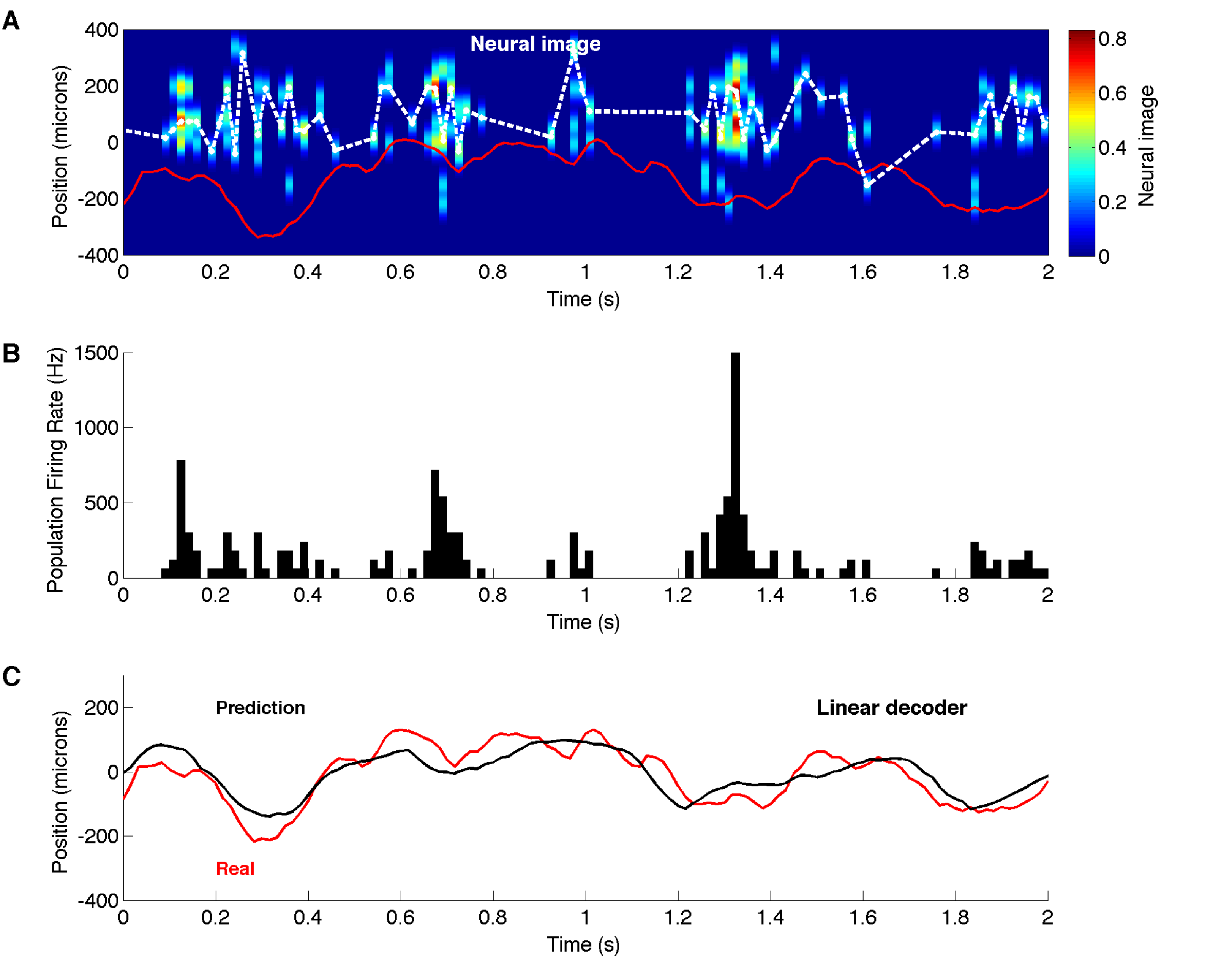}
\caption{ 
\textbf{Decoding based on the neural image. } 
\textbf{A:} Neural image in response to the moving bar. Color plot: neural image of the ganglion cell's population activity at each point in time. Black points: most likely position of the bar inferred from the peak in the neural image. Real trajectory in red.
\textbf{B:} Population firing rate summed over all the cells as a function of time, for the same time window than A. 
\textbf{C:} Prediction of the bar's trajectory using the linear decoding (black) ; real trajectory (red).
}
\label{Fig4}
\end{figure}

\subsection{Coding for motion in the receptive field surround}

From this lack of spatial structure we hypothesized that even the ganglion cells whose receptive fields were far from the bar carried information about the trajectory. 
To test this, we displayed the randomly moving bar stimulus in three different average locations, each separated by 430 microns and lasting 20 minutes (see Methods and fig.~5A). We then estimated the bar's trajectory for each stimulus ensemble one-at-a-time. The trajectory could be very well decoded for the three locations (fig.~5B, C, D). 

We plotted the individual decoding performance of different cells as a function of the distance between the cell's receptive field center coordinate and the mean position of the bar (fig.~5E). The decoding performance displayed a mild decrease as a function of distance (on average, a CC loss of $0.11$ per mm). This decrease was partly due to an increasing number of cells falling to $CC=0$. Many cells still carried information about the bar at distances of more than 800 $\mu$m away from the bar, while the average receptive field size is 115$\mu$m \cite{SegevTypes}. 
We then checked if these results could be explained by some ganglion cells having large receptive fields that would still overlap with the bar trajectory. We defined a normalized distance between the bar and each receptive field as the difference between their average position divided by the square-root of the product of their standard deviations (see methods). Many cells were still coding for the bar position even when this normalized distance was above 5, which corresponds to no overlap between the receptive field and the bar position distribution (fig.~5F). 
 These results show that the representation of the moving bar was distributed across a large population of cells spread widely in spatial location, far beyond what the extent of the receptive field center would predict. This also explains why the neural image failed to code for the bar position. 

\begin{figure}[!]
\centering
\includegraphics[width=0.9\textwidth]{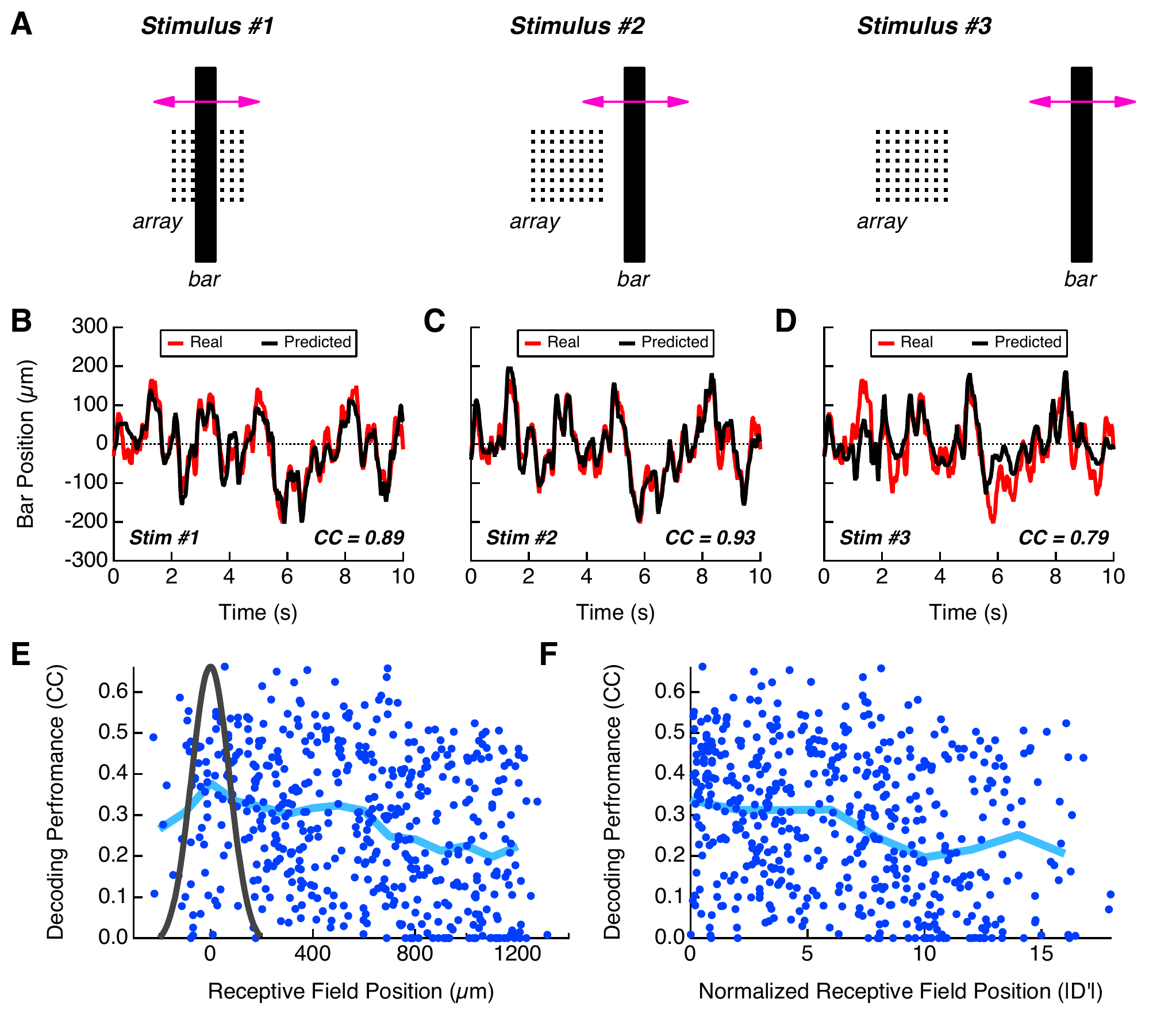}
\caption{ 
\textbf{Coding for motion in the receptive field surround. } 
\textbf{A:} Schematic of the experiment: the bar is randomly moved at three different average locations relative to the array. 
\textbf{B, C, D:} Prediction of the bar's trajectory using the linear decoding (black) ; real trajectory (red), for the three average locations above. 
\textbf{E:} Performance of linear decoding (blue) for individual ganglion cells (dots) plotted as a function of the distance between their receptive field center coordinate and the average bar position; probability distribution of bar position (black). Blue line: average decoding performance as a function of the distance.
\textbf{F:} Performance of linear decoding (blue) for individual ganglion cells (blue dots) plotted as a function of the normalized distance between their receptive field and the average bar position (see text). Blue line: average decoding performance as a function of the normalized distance. 
}
\label{Fig5}
\end{figure}

\subsection{A redundant and distributed code}

Because of the distributed nature of the code, we wondered how much redundancy was present in the ganglion cell population. As seen above (fig.~2), after a rapid initial increase, the decoding performance essentially saturated as a function of the number of cells. This saturation suggests that the information encoded by a new cell was highly redundant with the rest of the population. To quantify the redundancy in a principled way, we estimated the information rate between the real and predicted trajectories for single cells and for groups of cells of different sizes. 
Note that this measure is a lower bound on the true mutual information between the spike trains of a group of cells and the stimulus \cite{SpikesBook,BorstTheunissen}.

Information about the moving bar's trajectory encoded by subsets of ganglion cells increased as more cells were added (fig. 6A). But for the same number of cells, the performance varied substantially depending on the particular cells chosen in the subset. A natural hypothesis for this variability is that some cells carry more information about the stimulus than others. To explore this relationship, we compared the ``total'' information encoded by a group of ganglion cells (y-axis in fig.~6A and B) versus the sum of the individual informations encoded by each cell in the group (x-axis in fig.~6B). Plotted in these units, there was much less variability in the information encoded by different groups of ganglion cells. The same result was obtained in the guinea pig retina (fig.~6C,D). 

In making this comparison, we picked subsets randomly. To better test the predictive power of the sum of individual informations, we looked for the subsets with the best and the worst decoding performances for the same number of cells. There were too many combinations of cells having a given group size to allow for a comprehensive search over all possible subsets. Instead, we ordered the cells by their individual informations, and for each number of cells $N$, we estimated the total information for the $N$ ``best'' cells and the $N$ ``worst'' cells. While there was a large difference in the performance for the best and the worst subsets when plotted against the number of cells (fig.~6E), this difference was almost entirely compensated when plotting them against the sum of individual information (fig.~6F). For a group of ganglion cells having a similar sum of individual informations (i.e. a difference less than 0.5 bits$/s$), the average difference between worst and best subsets was 8\% of the best. Part of the remaining difference is likely due to the error in the estimation of the individual informations (horizontal error bars) due to the finite size of the data. In contrast, the ``worst'' cell groups had on average 55\% less information than "best" groups when matched for number of cells.

These results together show that we could predict the decoding performance of a group of ganglion cells from the properties of individual cells. 
In both the guinea pig and the salamander, small groups (less than 10 cells) had a total information that was close to the sum of individual informations, so the different cells carried nearly independent information. But as the sum of individual informations increased, it became much larger than the total information encoded by the group. Thus, the redundancy among ganglion cells (defined as total information divided by sum of individual informations) increased with the number of cells and became quite large as the performance saturated, with a 6.4-fold redundancy for the 123 cells in the salamander, and 6.6-fold for 140 cells in the guinea pig. 

\begin{figure}[!]
\centering
\includegraphics[width=0.9\textwidth]{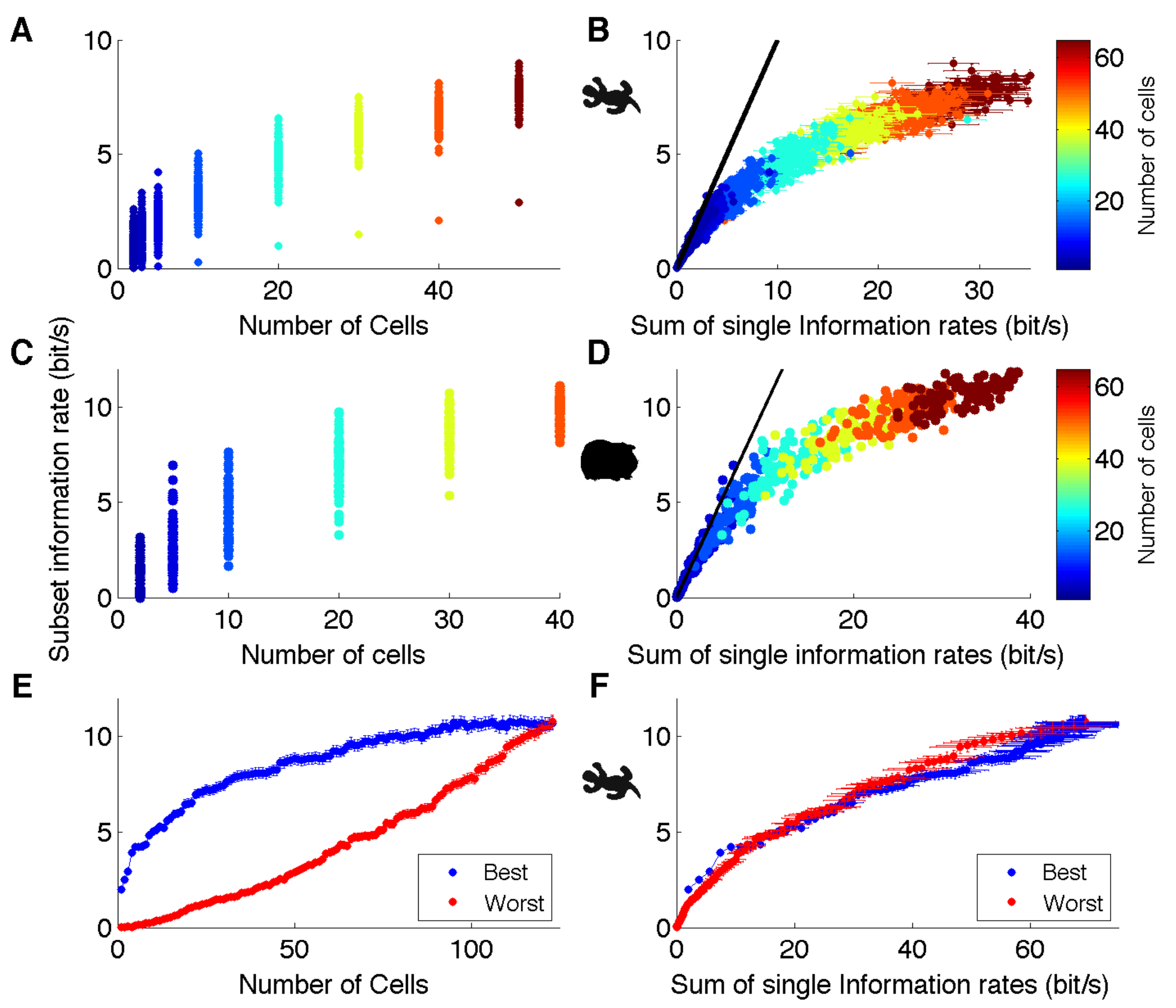}
\caption{ 
\textbf{Redundancy of the retinal code.}
\textbf{A, B:} Information rate obtained when decoding with different subsets of ganglion cells (dots) plotted against the number of cells (A) or the sum of the individual information rate of each cell of the subset (B); color scale indicates the number of cells in a subset. 
\textbf{C, D:} Plots analogous to panels A, B but for the guinea pig retina. 
\textbf{E, F:} Information rate for the best (blue) and worst (red) subsets of salamander ganglion cells (see text) plotted against the number of cells (E) or the sum of the individual information rate of each cell of the subset (F). 
}
\label{Fig6}
\end{figure}

This suggests that multiple subsets of cells might be able to encode the bar trajectory.  To find these disjoint subsets, we used L1 regularized decoding, a method that simultaneously minimizes the error in the prediction while setting to zero as many filter coefficients as possible (see Methods). Compared to the previous optimization where there were no constraints on the filter amplitude, here most filters were driven to zero (Fig 7A), but the prediction performance remained unchanged (CC$=$0.93). The filters with the highest amplitude corresponded to the neurons that were the most useful at decoding the trajectory. We ordered them from the highest to the lowest filter amplitude, and split them into groups of 10 that we used to decode the bar position. As expected, the cells with higher filter amplitude were better at decoding the trajectory than those with lower amplitude (fig. 7B), which confirmed that the L1 method picked the most informative cells. Of particular interest, we also found that the performance remained above CC $=$ 0.75 for the first seven groups, indicating many different groups of 10 neurons could decode the trajectory with high accuracy. We then looked for disjoint subsets that would give a prediction performance above $CC=$0.85, which was 90\% of the performance for the entire population. We first tried to decode using only the cell with the highest filter amplitude, and then kept adding cells to the subset, ordered from the highest to the lowest amplitude, until the decoding performance using the subset reached 0.85. We then restarted the same process after discarding the cells already used in the subset. By iterating this selection, we could find 6 disjoint subsets of increasing size that were all able to decode the trajectory of the bar (fig.~7C). These analyses together demonstrate that we could define 6 or 7 subsets with high decoding performance, similar to the estimated redundancy of information in the population.

\begin{figure}[!]
\centering
\includegraphics[width=1\textwidth]{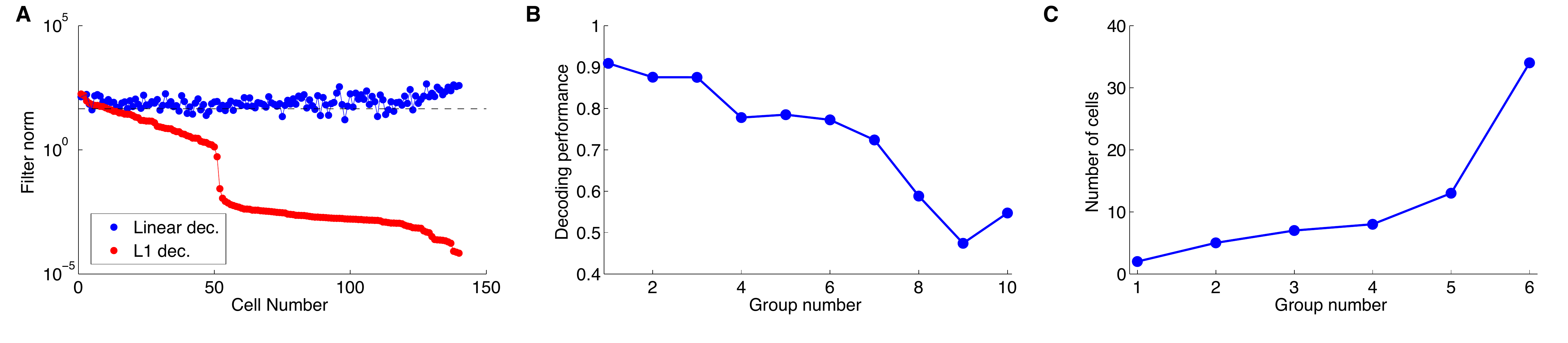}
\caption{ 
\textbf{Regularized decoding}
\textbf{A:} Comparison of the norm of the filters of each cell for the linear decoding (blue) against the regularized L1 decoding (red). The dashed line separate the 10 cells with the highest regularized norm. 
\textbf{B:} Decoding performance of groups of 10 cells sorted by their L1 norm, from the highest to the lowest. 
\textbf{C:} Number of cells in each subset that could reach a decoding performance of 0.85. 
}
\label{Fig7}
\end{figure}

\section{Discussion}

We have shown that we could reconstruct the entire time-varying trajectory of an object in complex motion from the activity of a large population of sensory neurons in the early visual system. Using linear decoding, populations of 100$+$ retinal ganglion cells were able to estimate the location of the object during its motion with an accuracy better than the spacing between cone photoreceptors in the salamander. Such hyperacuity performance has, to our knowledge, never before been demonstrated for a full time-varying reconstruction of a sensory stimulus.

The naive expectation was that many individual ganglion cells would fire spikes when an object was moving on their receptive field center and that the activity of the entire population would resemble a ``hill'' of higher activity that continuously tracked the location of the moving object. While the spatial map of the retina certainly localizes ganglion cell activity to some extent, our analysis revealed that neural activity was too sparse, and too distributed in space in a region extending into the receptive field surround, to closely track the object's location. Instead, the ganglion cells exhibited a highly distributed and redundant population code that allowed several distinct groups of cells to achieve super-resolution in tracking the location of a diffusively moving object.

\subsection{Comparison to Previous Results}

Linear decoding was first developed and applied to the H1 interneuron in the fly to estimate wide-field motion \cite{Bialek1991}. Linear decoding has also been applied to small populations of retinal ganglion cells to estimate the light intensity of a spatially uniform, white noise stimulus \cite{Warland1997,AlaRieke2011}, and to reconstruct a checkerboard stimulus for large populations of primate parasol cells \cite{pillow2008}. However, in these previous studies, the LN model worked well for the stimuli employed. Some of the most notable quantitative successes of the LN model have been demonstrated with spatially uniform stimulation \cite{Keat2001,Pillow2005,Fairhall2006}. Pillow et al (2008) showed that the LN model performed very well for parasol cells under their stimulus ensemble, perhaps due to the fact that the checkers were large enough to cover a cell's entire receptive field center. 

This is not the case for the kind of diffusive motion we used here, which includes pauses, starts, and reversals that trigger transient bursts of firing that cannot be explained by the LN model \cite{Schwartz2007,Chen2013}. One important factor in the failure of the LN model is the presence of nonlinear spatial subunits inside ganglion cell receptive field \cite{HochsteinShapley1976, Olveckzy2003,Baccus2008,Chen2013}. During spatially uniform stimulation, all nonlinear spatial subunits are activated together, allowing models that do not include this structure to approximate well the ganglion cell's firing rate. But during irregular motion, the subunits can be independently activated, such that a model without this structure performs poorly \cite{ChenPhD,Chen2013}. 

Other studies have tried to estimate a visual stimulus from the activity of populations of ganglion cells. The speed of a moving bar could be estimated from the activity of parasol ganglion cells \cite{FrechetteEJ}. However, this study was restricted to the case of motion at constant speed, so only one number was estimated: the speed of the moving bar. For the case of a moving texture that switched speeds every 500 ms, discrete classification of the speed was possible \cite{Thiel2007} (see also \cite{Greschner2002}). Similar classification studies have been performed in the visual cortex \cite{Radons1994, Oram1998}. Our study differs from these by reconstructing the full time-varying position of the bar, a task that is much more difficult than the classification of several stimuli or the estimation of a single parameter. 

\subsection{Structure of retina's population code }

One of the most significant surprises in this study was the lack of precise spatial structure of the retina's population code. 
One key factor was the complexity of our motion stimulus. Most early studies of how ganglion cells respond to motion involved an object moving at constant velocity \cite{Leonardo2013}, a case in which the picture of the neural image seems to hold. In our stimulus ensemble, the object moved at varied speed, including discontinuities of motion and moments of high acceleration. Motion discontinuities like the sudden onset or reversal of motion trigger transient bursts of firing \cite{Schwartz2007,Chen2013}, as does acceleration in general \cite{Thiel2007}. 

Another striking result was how well ganglion cells encoded diffusive motion of a dark bar in their surrounds, which partly explained the diffuse spatial structure of the code. Motion in the surround has long been known to be able to generate excitatory responses in ganglion cells \cite{McIlwain1964}. Our analysis demonstrates that this activation is not just a global alert signal, but can be used to encode the precise trajectory of the bar. Detailed investigation has revealed that the response of ganglion cells to gratings drifting in their surround can be modeled as the pooling of excitatory and inhibitory non-linear subunits \cite{Passaglia2001, Passaglia2009}. Many other studies have demonstrated the presence of disinhibitory circuits within the receptive field center, involving transient amacrine cells turning off sustained inhibition from sustained amacrine cells \cite{Taylor2010,RusselWerblin2010,ManuBaccus2012}. These circuits may allow precise information about motion in the surround to be conveyed by the firing of ganglion cells.

\subsection{A flexible code}

Clearly, the distributed and redundant neural code that we have observed is not how a human engineer would design a system to track motion. In most theoretical studies where a population of neurons has to code for the value of a parameter, the single cell selectivity is modeled by a tuning curve. As a result, the activity in response to the stimulus is local, and neurons code best for stimuli at the peak of their sensitivity, or at the side of their tuning curves, depending on the noise level \cite{GoldmanButts2006}. Experimentally, this corresponds to the case of the neural image. While the peak of neural activity does track objects moving at constant velocity \cite{Berry1999,Leonardo2013}, we have shown that, in our case of diffusive motion, the retinal code was too sparse and spatially diffuse for the peak of neural activity to track the position of the bar. 

Even though the retinal code did not have a straightforward spatial organization, it could still be read out with one of the simplest of all decoding mechanisms: the linear decoder. One of the implications of the success of the linear decoder is that it makes information easy to extract by subsequent neural circuits \cite{GollischMeister2010,RigottiFusi2010}. More generally, a fundamental function of sensory circuits might be to compute and actively maintain an ``explicit'' or linearly decodable representation of the most relevant features of the environment \cite{DiCarloCox2007}. This appears to be the case for IT cortex: the identity of an object can be decoded linearly from IT neurons, but not from V1 neurons \cite{DiCarloReadout2005}. Our results thus show that in this sense, the retina has an ``explicit'' representation of the position of a moving object.  

The high redundancy of the retinal code was both notable and somewhat surprising. Why might such a distributed and redundant code be a beneficial organization? One advantage is that neural circuits in the central brain would have access to high precision motion tracking information from several groups of ganglion cells. This is useful because there are multiple features in the visual scene beyond just the location of a moving object that the brain might want to follow. For instance, if such circuits sampled all $\sim$100 ganglion cells with receptive fields overlapping one point in visual space, they could extract high resolution spatial and chromatic information about object identity in addition to the object's trajectory. Alternatively, if $\sim$100 ganglion cells with distributed spatial locations were sampled, then information about the characteristics of a larger object could be extracted in addition to its motion trajectory. Thus, the distributed, redundant population code maintains maximum flexibility with respect to the purposes of downstream neural circuits. 

Of course, our analysis does not establish whether the brain uses a linear decoder nor how it might learn effective decoding kernels. In particular, we have shown that the decoders that reached the best performance had to be learned on the moving bar data directly. This means that the detailed parameters of the decoder that achieve high performance might depend on the properties of the stimulus ensemble. 

By demonstrating the broad spatial selectivity of ganglion cells, our study shows that neurons with complex tuning curves or mixed selectivity \cite{RigottiFusi2013} do not appear only when merging information from different modalities in associative areas of the brain, but already at the earliest stage of sensory processing. Interestingly, these features of the retinal code were only apparent when the stimulus dynamics were sufficiently rich. The redundancy of trajectory representation that we uncovered could be a trade-off between discrimination and generalisation \cite{Barak2013}. Future studies will have to understand how a representation of motion that is invariant to the context can be extracted from the retinal activity.

\newpage

\section{Methods}

\subsection{Recordings}

Retinal tissue was obtained from larval tiger salamanders (Ambystoma tigrinum) of either sex and continuously perfused with oxygenated Ringer's medium at room temperature. For guinea pig experiments, the tissue was perfused with AMES solution and maintained at 37~$^{o}$C. Ganglion cell spikes were recorded extracellularly from a multi-electrode array with 252 electrodes spaced 30 $\mu$m apart (custom fabrication by Innovative Micro Technologies, Santa Barbara, CA). Details of the recording and spike sorting methods are described elsewhere (\cite{SegevMethod,Method252}. Experiments were performed in accordance with institutional animal care standards. 

\subsection{Visual stimulation}

The stimuli were displayed on a CRT screen with a 60 Hz refresh rate \cite{PuchallaBerry}. The stimulus presented was a dark bar of 100\% contrast moving randomly over a gray background. The trajectory was given by Brownian motion with a spring-like force to bring the bar back to the center: 
\begin{equation}
\small
\frac{dv}{dt} = -\frac{v}{\tau} + \sigma\Gamma(t) - \omega_0^2 x
\label{eq:stimulus}
\end{equation}
where $x$ is the position of the bar, $v=\frac{dx}{dt}$ is the velocity of the bar, $\sigma^2=0.05$, $\Gamma(t)$ is a Gaussian white noise, $\omega_0=9.42$ and $\tau = 50$~ms. 

For the experiments where we tested the decoding performance as a function of frequency, the trajectory was sampled every stimulus frame (16.7 ms) from a Gaussian distribution of standard deviation 60 $\mu$m and low-pass filtered at a frequency of 5 Hz. 

Spatio-temporal receptive fields were measured by reverse correlation to a flickering checkerboard composed of squares of 69 $\mu$m size that were randomly selected to be black or white at a rate of 30 Hz \cite{SegevTypes}.

Since salamander eyes are $\sim$4 mm in diameter, one degree of visual angle should correspond to approximately 35 microns in the retina plane, assuming a focal length of 2 mm. So the size of the array corresponds to 12-13 degree of visual angle. 

\subsection{Linear decoding}

We used a linear model that takes the spike trains as an input and gives a prediction about the position of the bar as an output:
\begin{equation}
\small
p(t) = \sum_{i,j} K_i(t-t_{ij}) + C
\end{equation}
where $t_{ij}$ is the $j$-th spike of the neuron $i$, and $p(t)$ is the predicted position of the bar over time. The filters $K_i$ and the constant $C$ are found by minimization of $\langle (p(t) - x(t))^2 \rangle$ (least square minimization) \cite{Warland1997}, where $x(t)$ is the real position of the bar, and angular brackets denote averages over time. The filters extended 500 ms before and after the spike. 

All the results shown here were cross-validated: we trained our model on $2/3$ of the data, and tested it on the other $1/3$. We used one hour of recordings, which corresponds to 40 minutes for the training set, and 20 minutes for the test set. 

To characterize decoding performance, we estimated the normalized cross-correlation (CC) between the prediction and the real trajectory. For two signals $x(t)$ and $y(t)$, it is classically defined as $CC = \frac{\langle 
(x(t) - m_x).(y(t) - m_y) \rangle }{\sigma_x.\sigma_y}$, where $m_x$ and $\sigma_x$ are the mean and standard deviation of $x(t)$, respectively, and $\langle \rangle$ denotes an average over time. 

\subsection{Position-Frequency analysis}

To get an estimation of the error signal as a function of both the real position and the frequency, we filtered the error $e(t) = p(t) - x(t)$ by a bandpass filter that did not introduce a phase delay, i.e. a Morlet Wavelet. The wavelet was normalized so that the averaged squared value of the filtered signal matched the corresponding value in the total power spectrum. 
For each possible position, we picked the times where the bar was at this position, and took the average of the corresponding squared values in the filtered signal. Formally, if $e_f(t)$ is the filtered and squared version of $e(t)$ for frequency $f$, then the error $e(f,p)$ for position $p$ and frequency $f$ was computed as: 
\begin{equation}
\small
\frac{1}{N_p} \sum_{i=1}^{N_p} e_f(t_i) 
\end{equation}
where the $t_i$'s are all the times such that $x(t_i)=p$, and $N_p$ is the number of such points. 

\subsection{Neural image}

To construct the neural image, we assigned a spatial position to each cell as the peak of its receptive field. The activity of the cells was binned in 16.6 ms bins (corresponding to the refresh rate of the stimulus). At each time bin, we counted the number of spikes emitted by the cells at each spatial location. The resulting matrix was then smoothed across spatial positions with a Gaussian smoothing kernel of width 21 $\mu$m. The ``most likely'' position was obtained by taking the peak location of the neural image at each time where there was at least one cell firing.

\subsection{Information estimation}

The information rate between the true and decoded bar trajectory was estimated from their mutual coherence, $\gamma(f)$, as $I=-\int_0^{f_{\rm max}} df\;\log_2(1-\gamma^2(f))$ \cite{BorstTheunissen}. Debiased coherence was computed using Chronux \cite{MitraBokil} using trajectory windows of duration $256 / 60 s$ sampled at 60 Hz. The frequency range of integration $[0,f_{\rm max}]$ was determined as the contiguous range where the estimated coherence is above zero at $10^{-2}$ significance level. Alternatively we considered a larger frequency range by lowering the significance threshold to $10^{-1}$, and performing explicit debiasing by estimating the information on multiple subsets of the data of various sizes, and extrapolating to infinite sample size \cite{Strong1998}. The results agreed within error bars. To validate our method, we generated synthetic ``real'' and ``decoded'' traces with Gaussian statistics and power spectra that were similar to the real traces, and of similar length. In this case the true information is analytically computable. We compared this exact result to our estimators to assess their accuracy and verify that the chosen parameters (window length, determination of frequency range) were suitable. For information rates above 0.5 bit$/$s the estimated error in the information rate is between 5-10\%, while for rates of $\sim$0.1 bit$/$s the error increases to about 20\%. 

The relation between the normalized cross-correlation and the mutual information estimated this way was examined. If there were no temporal correlation, the mutual information would be 
$I = -0.5 \log{(1 - CC^2)}$. 
The measured information was significantly different from the value predicted by this formula. This is unsurprising since the formula does not take into account the different frequencies. However, we could fit the information values with a linear interpolation of this formula, i.e. $I = -0.5 \alpha \log{(1 - CC^2)} + \beta$. With this formula, we could explain 98\% of the variability in the information values. So an empirical relationship could be inferred between the two quantities. Note that $\alpha$ was larger than one, indicating a synergy between frequencies. 
Redundancy in a subset of cells $A$ was defined as $R = 1 - \frac{MI_{A}}{\sum_{i \in A} M_{i}}$, where $MI_{A}$ is the mutual information for the subset, and $M_{i}$ the mutual information for each cell. 

\subsection{Regularized decoding}

In L1-regularized decoding the least squares minimization problem $\langle (p(t)-x(t))^2\rangle$ is substituted by $\langle (p(t)-x(t))^2\rangle+\lambda\sum_i \| K_i(t)\|_1$ where $\lambda\geq 0$ is the regularization parameter. Unlike the simple least squares case, this regularized minimization cannot be reduced to a simple straightforward linear algebra problem and a numerical solution is required. For this purpose we have used the implementation of the interior-point method by Boyd et al \cite{Kim2007}. 
The value of the regularization parameter $\lambda$ is usually chosen to minimize the mean-squared error (MSE). However, our goal is to assign zero weight filters to as many cells as possible while keeping a good decoding performance. Therefore, we choose a larger-than-optimal regularization parameter by allowing a 5$\%$ decrease in performance (CC) and a 20$\%$ increase of the MSE. Also, we run a 5-fold cross-validation procedure on the training set to validate the choice of regularization parameter.

\subsection{Normalized distance}

For each cell $i$, we estimated its receptive field using a classical checkerboard stimulus (see above) and convolved it with the bar. From the resulting spatial distribution, we estimated the average position $m_i$ and the standard deviation $s_i$. From the distribution of the bar position, we also estimated the mean position $m_b$ and the standard deviation $s_b$. The normalized distance was then defined as $d_i = \frac{|m_b - m_i|}{\sqrt{s_b s_i}}$

\section{Acknowledgments}

We thank Stephanie Palmer for help with the stimulus design, William Bialek, Jonathan Pillow and Maureen Clerc for helpful suggestions, Eric Chen for help with the experiments. This work was supported by grants EY 014196 and EY 017934 to M.J.B, and ANR OPTIMA to O.M., and by the French State program ``Investissements d'Avenir'' managed by the Agence Nationale de la Recherche [LIFESENSES: ANR-10-LABX-65], the Austrian Research Foundation FWF P25651 to VBS and GT. V.B.S. is partially supported by contract MEC, Spain (Grant No. AYA2010- 22111-C03-02 and FEDER Funds). The funders had no role in study design, data collection and analysis, decision to publish, or preparation of the manuscript.

\clearpage

\end{document}